\documentclass[twocolumn,showpacs,preprintnumbers,superscriptaddress,amsmath,amssymb,prl]{revtex4}

\usepackage{graphicx}
\usepackage{dcolumn}
\usepackage{bm}
\usepackage{quantum}

\begin{document}

\preprint{APS/123-QED}

\title{Generation of photon number states}
\author{Edo Waks}
\affiliation{Quantum Entanglement Project, ICORP, JST, E.L.
Ginzton Laboratories, Stanford University, Stanford, CA 94305 }
\author{Eleni Diamanti}
\affiliation{Quantum Entanglement Project, ICORP, JST, E.L.
Ginzton Laboratories, Stanford University, Stanford, CA 94305 }
\author{Yoshihisa Yamamoto}
\affiliation{Quantum Entanglement Project, ICORP, JST, E.L.
Ginzton Laboratories, Stanford University, Stanford, CA 94305 }
\affiliation{NTT basic research, Atsugi, Kanagawa, Japan}

\date{\today}

\begin{abstract}

The Visible Light Photon Counter (VLPC) has the capability to
discriminate photon number states, in contrast to conventional
photon counters which can only detect the presence or absence of
photons.  We use this capability, along with the process of
parametric down-conversion, to generate photon number states. We
experimentally demonstrate generation of states containing 1,2,3
and 4 photons with high fidelity.  We then explore the effect the
detection efficiency of the VLPC has on the generation rate and
fidelity of the created states.

\end{abstract}

\maketitle

Photon number states play an important role in quantum optics.
Such states exhibit effects which contradict classical
electromagnetic theory, such as anti-bunching and negativity of
the Wigner function~\cite{WallsMilburn95}.  They also have
important practical applications.  In optical telecommunication,
they achieve an optimal channel
capacity~\cite{YamamotoHaus86,CavesDrummond94}. They can also
improve the sensitivity of an interferometer to the Heisenberg
limit~\cite{HollandBurnett93}.  In this limit, the minimum
detectable phase shift is inversely proportional to the number of
photons, rather then the square root of the number achieved by the
standard quantum limit.

Recently, there has been tremendous effort in generating single
photon states.  This is often achieved by using single emitters
such as molecules, quantum dots, or diamond color
centers~\cite{SantoriPelton01,KimBenson99,LounisMoerner00,Michler00,Moreau01,Beveratos02,Yuan02}.
Although these sources hold great promise, they currently suffer
from substantial collection losses. This creates a large vacuum
contribution, which means that the generated state greatly
deviates from one with exactly one photon. Furthermore, the
extension of these devices to generating higher order photon
number states remains a difficult problem.  The generation of a
two photon number state has been recently demonstrated in the
microwave regime~\cite{VarcoeBrattke01}, using Rydberg atoms in a
high-Q cavity. There have also been proposals for generation
higher order optical photon number states using trapped atoms in
an optical cavity~\cite{BrownDani03}. The implementation of these
schemes remains experimentally challenging.

There are generally two ways to engineer the quantum state of a
system.  The first is to start the system in a well known initial
state, and turn on a unitary time evolution which transforms the
initial state to the desired state.  The second method is to make
a non-destructive measurement of the quantum system, and project
it onto the desired Hilbert space using the projection postulate
of quantum mechanics.  In this letter, we use the latter method of
projection and conditional post-selection to generate a photon
number state.  This is done using the non-linear process of
parametric down-conversion (PDC) in conjunction with a photon
number detector known as the Visible Light Photon Counter (VLPC).
In PDC, a bright pump field is injected into a material which
exhibits a second order non-linear dipole response.  This
non-linearity can cause a pump photon to spontaneously split into
two photons of lower energy, referred to as the signal and idler.
By appropriately adjusting the phase matching conditions of the
process, the signal and idler can be made to propagate in
different directions, allowing them to be spatially resolved.
Since these two photons always come in pairs, if  a signal photon
is detected, there must be an idler photon in the conjugate mode.
If a photon counter is placed in front of the signal arm and
detects one photon, the corresponding idler arm is prepared in a
state containing only one photon by the projection postulate of
quantum mechanics. The photon counter in essence performs a
non-destructive measurement of the idler arm.

The above scheme can be extended to generation of higher order
number states.  Suppose a short pump laser pulse is injected into
a non-linear crystal.  We can define the photon number operator
$\n = \int_T \adag\left( t \right) \a \left( t \right) dt$, where
the integral is taken over the pulse duration of the pump and $\a
\left( t \right)$ is the bosonic photon annihilation operator in
the time domain.  If n signal photons are detected by a photon
counter in a given pulse, then the idler pulse is projected onto a
state $\|\psi_i>$ which satisfies the condition
  \begin{equation}
    \n \|\psi_i> = n \|\psi_i>.
  \end{equation}
That is, the idler mode is an eigenstate of the number operator
with eigenvalue $n$.  States which satisfy this special property
are known as photon number states.

In order to implement the above scheme, the photon counter must be
able to determine the exact number of photons in the signal arm
within the short time duration of the pump pulse. Conventional
photon counters such as avalanche photo-diodes (APDs) cannot do
this, because they suffer from long dead time and large
multiplication noise.  Such detectors can only distinguish the
zero photon case from the non-zero photon case in a pulse, making
it impossible to generate photon number states.

Recently, a new type of photon counter known as the Visible Light
Photon Counter (VLPC)~\cite{TurnerStapelbroek93}, has been shown
to have the capability to distinguish different photon number
states with high quantum
efficiency~\cite{TakeuchiKim99,Atac94,KimTakeuchi99,WaksInoue03}.
When two photons are simultaneously detected by the VLPC, it
generates an electrical pulse which is twice as large. This
pattern holds for higher photon numbers.  The pulse height can
then be used to determine the number of detected photons.

The VLPC can do photon number detection due to two unique
properties.  First, the VLPC has an active area of 1mm in
diameter.  When one photon is detected, it forms a 5$\mu$m
diameter dead spot on the detector surface, leaving the rest of
the active area available for subsequent detection events.  As
long as the probability that two photons land on the same spot is
small, which is true if the incident light in not too tightly
focussed, all of the photons will be detected.  This circumvents
the dead time problem which conventional APDs suffer from. Second,
the VLPC has extremely low multiplication noise. Multiplication
noise refers to the fluctuations in the number of electrons
emitted by a detector in response to a photodetection event. These
fluctuations limit the ability of the detector to infer photon
number~\cite{WaksInoue03}. Multiplication noise is quantified by a
parameter known as the Excess Noise Factor $F=\langle
M^2\rangle/\langle M \rangle^2$, where $M$ is the number of
electrons emitted by the detector due to a photon detection. In
the ideal limit where $F=1$, the detector emits a deterministic
number of electrons for each detection.  APDs with single carrier
multiplication, which achieves the best noise performance, still
have a large excess noise factor of $F=2$~\cite{McIntyre66},
making it impossible to discriminate photon numbers.  The VLPC, in
contrast, achieves nearly noise free multiplication with
$F=1.03$~\cite{KimYamamoto97,WaksInoue03}.

\begin{figure}
\centerline{\scalebox{0.6}{\includegraphics{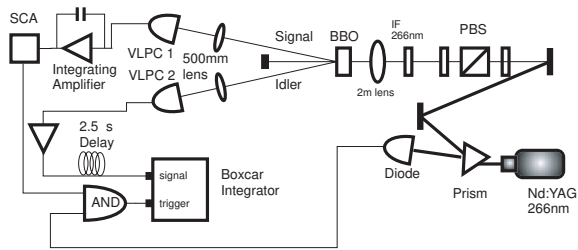}}}
\caption{Experimental setup for photon number generation.}
\label{Fig:ExpSetup}
\end{figure}

The experimental setup for generating photon number states is
shown in Figure~\ref{Fig:ExpSetup}.  A 266nm pump source is
generated from the fourth harmonic of a Q-switched Nd:YAG laser.
The pump pulses have a duration of 20ns, and a repetition rate of
45kHz.  A dispersion prism separates the fourth harmonic from the
residual second harmonic, which is used to illuminate a high speed
photodiode to generate a triggering signal.  The fourth harmonic
pumps a BBO crystal, which is set for non-collinear degenerate
phase-matching. In this condition, the signal and idler waves are
both 532nm in wavelength and have a divergence angle of 1 degree
from the pump. The pump is loosely focused before the BBO crystal
to achieve a minimum waste at the collection lens.  This results
in a sharper two-photon image which enhances the collection
efficiency~\cite{MonkenSoutoRibeiro98}.

Two VLPC detectors are used in this experiment.  Each detector is
held in a separate helium bath cryostat and cooled down to 6-7K,
which is the optimum operating temperature.  The VLPC is sensitive
to photons with wavelengths of up to 30$\mu$m, so it must be
shielded from room temperature thermal radiation.  This is
achieved by encasing the detector in a copper shield, which is
cooled down to 6K.  Acrylic windows at the front of the copper
shield are used as infra red filters.  These windows are highly
transparent at visible wavelengths and simultaneously nearly
opaque at 2-30$\mu$m wavelengths.

\begin{figure}
\centerline{\scalebox{0.7}{\includegraphics{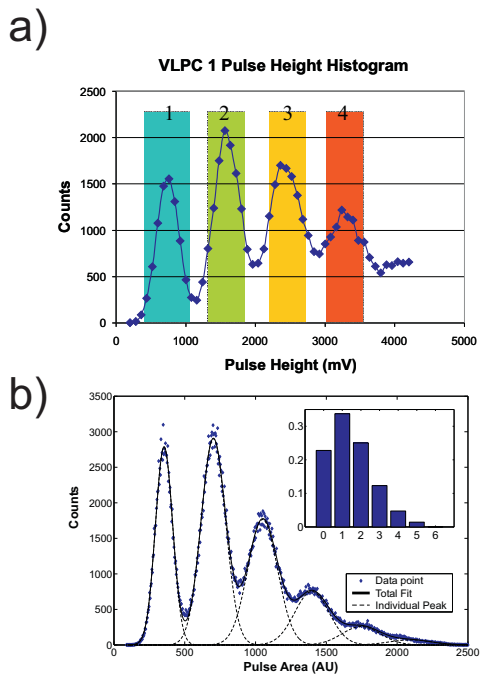}}}
\caption{Photon number detection from the VLPCs. {\bf a}, pulse
height histogram from VLPC 1 after an integrating amplifier.  The
histogram features a series of peaks corresponding to the photon
number distribution of the parametric down-conversion output.  An
SCA is used four different regions, corresponding to the different
photon number events, shown by the shaded regions. {\bf b}, a
pulse area histogram of VLPC 2 when no post-selection is done from
VLPC 1. Each peak is fit to a gaussian to calculate the photon
number distribution, shown in the inset.} \label{Fig:PulseHeight}
\end{figure}

VLPC 1 is used as the triggering detector which detects the number
of photons generated in the signal arm on a given laser pulse. The
output of VLPC 1 is amplified by an integrating amplifier, which
generates an electrical pulse whose height is proportional to the
number of emitted electrons.  The height of the pulse is
discriminated by a single channel analyzer (SCA).  A logical AND
is performed between the output of the SCA and the output of the
photodiode to reject all detection events which occur outside of
the pulse duration.  Figure~\ref{Fig:PulseHeight}a shows a pulse
height histogram of VLPC 1.  This histogram features a series of
peaks corresponding to different photon number states.  The SCA
can select pulse heights corresponding to one, two, three, and
four photon events. The decision window set by the SCA for each
photon number event is shown by the shaded areas in the figure.

VLPC 2 is placed in the idler arm, and used to verify that the
correct photon number state was generated.  The output of VLPC 2
is amplified, and connected to the signal input of a boxcar
integrator.  On each pulse from the SCA, the output of VLPC 2 is
integrated over a 2$\mu$s window, which is sufficiently large to
encompass the entire electrical pulse (determined by the bandwidth
of subsequent amplifiers).   Figure~\ref{Fig:PulseHeight}b shows a
pulse area distribution of VLPC 2 with no post-selection from VLPC
1, at a pump excitation power of 40$\mu$W.  This distribution also
features a series of peaks corresponding to the different photon
numbers, starting with the first peak which is a zero photon
event. In order to calculate the photon number distribution, each
peak is fit to a gaussian. The area of each gaussian is normalized
by the total area of all the peaks. The calculated photon number
distribution is shown in the inset.

Both VLPCs have imperfect detection efficiencies due to internal
detection losses of the VLPC itself, as well as external losses
from the collection optics.   The efficiency is measured by
comparing the coincidence rate between the two detectors to the
singles count rates. From this measurement we determine the
detection efficiency of VLPC 1 to be 0.68, and VLPC 2 to be 0.58.

The detection efficiency of the monitoring detector (VLPC 2) can
be corrected for, because we are interested mainly in how many
photons were actually present in the idler arm, not how many have
been detected.  To do this, we can define $p_j$ as the probability
that j photons were generated in the idler pulse, and $f_i$ as the
probability that i photons were detected.  These two probabilities
are related by the linear loss model $f_i = \sum_{j=i}^{\infty}
\binom{j}{i} \eta^i \left( 1 - \eta \right)^{j-i} p_i$. The above
transformation needs to be inverted.  In order to do this we
truncate the photon number distribution at some number $n$, which
is sufficiently large so that $p(n+1)\approx 0$ is a good
approximation.  The initial and final probability distributions
are then related by a constant matrix, which can simply be
inverted.  This matrix can also be modified to account for dark
counts in a straightforward way.

The detection efficiency of the triggering detector (VLPC 1) plays
a more subtle role in the experiment.  Detection losses can result
in a higher order photon number state being misinterpreted by the
detector as the correct photon number.  Because of this, the
probability distribution in the idler arm will no longer be an
exact photon number, but a mixture of the desired number plus
higher order number states.  This results in a degradation of the
fidelity, defined as the overlap between the desired state and the
actual generated state.  For a state containing $n$ photons the
fidelity is simply the probability that $n$ photons are generated.

\begin{figure}
\centerline{\scalebox{0.55}{\includegraphics{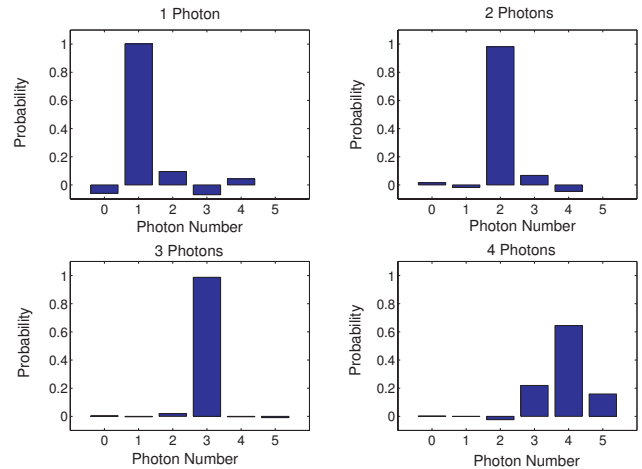}}}
\caption{Results of the photon number generation experiment.  The
photon number distributions, after correcting for detection
efficiency and dark counts of VLPC 2, are plotted for the reported
number n=1,2,3, and 4 by VLPC 1.  The detection efficiency and
dark counts of VLPC 2 are measured independently.}
\label{Fig:LowPowerDist}
\end{figure}

When the pumping intensity is low, the efficiency of the
triggering detector does not play a major role.  If the triggering
detector sees $n$ photons, it is true with very high probability
that the same number of photons are present in both arms. The
probability that there were more photons present is extremely low,
because low pumping intensity makes the probability of generating
higher order photon pairs negligible.
Figure~\ref{Fig:LowPowerDist} shows the experimental results at
this low pumping intensity regime. The figure shows the photon
number distributions measured by VLPC 2, after correcting for
detection efficiency and dark counts, when VLPC 1 post-selects a
one, two, three, and four photon event. For one, two, and three
photon post-selection, a nearly ideal photon number state is
generated. For four photon post-selection however, there are
contributions from three and five photon number states. These
contributions are attributed to the smearing between the four
photon peak and its next nearest neighbors in the pulse height
histogram of VLPC 1 (Fig. ~\ref{Fig:PulseHeight}a).  The smearing
is caused by buildup of multiplication noise, which puts a limit
on the photon number resolution at higher
numbers~\cite{WaksInoue03}. Note that, in a few cases, the
corrected probability distribution becomes negative. This
erroneous effect is caused by numerical errors in the probability
reconstruction.

\begin{figure}
\centerline{\scalebox{0.45}{\includegraphics{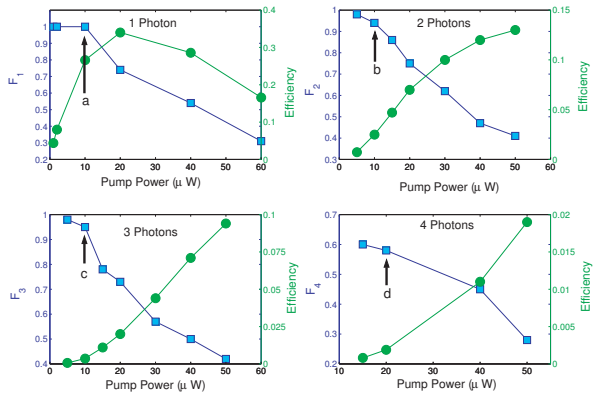}}}
\caption{The generation rate (circles) and fidelity (squares) are
plotted as a function of pump power.  The fidelity corresponds to
the left y-axis, while the generation efficiency corresponds to
the right y-axis.} \label{Fig:FidelityVsRate}
\end{figure}

As we increase the pumping power, the imperfect detection
efficiency of VLPC 1 will result in degraded fidelity, as was
previously discussed. On the other hand, higher pump powers will
increase the probability that the correct number of photons were
generated, and hence the generation rate of the desired photon
number state. This leads to a natural tradeoff between the state
fidelity and generation rate of photon number states.  The extent
of the tradeoff is determined mainly by the detection efficiency
of VLPC 1. In order to determine this tradeoff, we measure the
fidelity and generation rate as a function of pump power for the
four different photon number states.  The results are shown in
Figure~\ref{Fig:FidelityVsRate}. For all four cases, increasing
the pump power results in higher generation rates but decreased
fidelity. In the figure, four data points are labelled as a, b, c,
and d. These points designate the pumping intensity before the
fidelity begins to drop for one, two, three, and four photon
number states. The generation rate and fidelity at these four
points are: a - 11800 Hz and $F_1$=1.0, b -1100 Hz and $F_2$=0.94,
c - 160 Hz and $F_3$=0.95, d - 84 Hz and $F_4$=0.6. The above
generation rates are already large enough to be useful for
practical experiments. These rates are achieved using only a 45kHz
pump repetition rate. The rates could be significantly enhanced by
simply using a pump laser with a faster repetition rate.

We conclude by discussing some of the future extensions of this
work.  There are two main drawbacks to the experimental scheme we
present in this letter.  First, the setup does not allow for
generation of photon number states on demand.  The time in which
the number state is generated is known but uncontrollable.  This
problem can be solved in principle by setting up a large parallel
array of photon number generators along with an optical switch. By
making the array large, the probability that at least one of the
number state generators will generate the right photon number can
be made arbitrarily close to unity.  Of course, with the current
setup such a solution is impractical, but perhaps the engineering
of VLPC arrays, along with more compact two photon sources based
on waveguide technology, may make such a solution possible.

The second drawback of our experiment is that the pump pulse to
the down-conversion crystal has a 20ns duration, as opposed to the
coherence time of the down-converted photons which is on the order
of femtoseconds.  This means that, although the states we generate
are eigenfunctions of the photon number operator, the quantum
mechanical wavefunctions of the photons is not fixed. In fact the
wavefunction of each photon will generally fluctuate from shot to
shot.  A more ideal source would generate a state in which both
the number and wavefunction of the photons is constant. The best
case scenario would be the generation of a so-called n-photon Fock
state, which is a state containing n-photons with identical
wavefunctions.  Our setup could be extended to generate such a
state by replacing the pumping laser with a femtosecond pulse
laser, so that the pulse duration is on the order of the coherence
length of the photons. This could open up the door for interesting
new experiments in quantum information using photon number states.

  \bibliographystyle{apsrev}
  \bibliography{MultiPhotonEps0808}

\begin{thebibliography}{21}
\expandafter\ifx\csname natexlab\endcsname\relax\def\natexlab#1{#1}\fi
\expandafter\ifx\csname bibnamefont\endcsname\relax
  \def\bibnamefont#1{#1}\fi
\expandafter\ifx\csname bibfnamefont\endcsname\relax
  \def\bibfnamefont#1{#1}\fi
\expandafter\ifx\csname citenamefont\endcsname\relax
  \def\citenamefont#1{#1}\fi
\expandafter\ifx\csname url\endcsname\relax
  \def\url#1{\texttt{#1}}\fi
\expandafter\ifx\csname urlprefix\endcsname\relax\def\urlprefix{URL }\fi
\providecommand{\bibinfo}[2]{#2}
\providecommand{\eprint}[2][]{\url{#2}}

\bibitem[{\citenamefont{Walls and Milburn}(1995)}]{WallsMilburn95}
\bibinfo{author}{\bibfnamefont{D.}~\bibnamefont{Walls}} \bibnamefont{and}
  \bibinfo{author}{\bibfnamefont{G.}~\bibnamefont{Milburn}},
  \emph{\bibinfo{title}{Quantum optics}} (\bibinfo{publisher}{Springer},
  \bibinfo{address}{New York}, \bibinfo{year}{1995}).

\bibitem[{\citenamefont{Yamamoto and Haus}(1986)}]{YamamotoHaus86}
\bibinfo{author}{\bibfnamefont{Y.}~\bibnamefont{Yamamoto}} \bibnamefont{and}
  \bibinfo{author}{\bibfnamefont{H.}~\bibnamefont{Haus}},
  \bibinfo{journal}{Rev. Mod. Phys.} \textbf{\bibinfo{volume}{54}},
  \bibinfo{pages}{1001} (\bibinfo{year}{1986}).

\bibitem[{\citenamefont{Caves and Drummond}(1994)}]{CavesDrummond94}
\bibinfo{author}{\bibfnamefont{C.}~\bibnamefont{Caves}} \bibnamefont{and}
  \bibinfo{author}{\bibfnamefont{P.}~\bibnamefont{Drummond}},
  \bibinfo{journal}{Rev. Mod. Phys.} \textbf{\bibinfo{volume}{66}},
  \bibinfo{pages}{481} (\bibinfo{year}{1994}).

\bibitem[{\citenamefont{Holland and Burnett}(2001)}]{HollandBurnett93}
\bibinfo{author}{\bibfnamefont{M.}~\bibnamefont{Holland}} \bibnamefont{and}
  \bibinfo{author}{\bibfnamefont{K.}~\bibnamefont{Burnett}},
  \bibinfo{journal}{Phys. Rev. Lett.} \textbf{\bibinfo{volume}{86}},
  \bibinfo{pages}{1502} (\bibinfo{year}{2001}).

\bibitem[{\citenamefont{Santori et~al.}(2001)}]{SantoriPelton01}
\bibinfo{author}{\bibfnamefont{C.}~\bibnamefont{Santori}} \bibnamefont{et~al.},
  \bibinfo{journal}{Phys. Rev. Lett.} \textbf{\bibinfo{volume}{86}},
  \bibinfo{pages}{1502} (\bibinfo{year}{2001}).

\bibitem[{\citenamefont{Kim et~al.}(1999{\natexlab{a}})\citenamefont{Kim,
  Benson, Kan, and Yamamoto}}]{KimBenson99}
\bibinfo{author}{\bibfnamefont{J.}~\bibnamefont{Kim}},
  \bibinfo{author}{\bibfnamefont{O.}~\bibnamefont{Benson}},
  \bibinfo{author}{\bibfnamefont{H.}~\bibnamefont{Kan}}, \bibnamefont{and}
  \bibinfo{author}{\bibfnamefont{Y.}~\bibnamefont{Yamamoto}},
  \bibinfo{journal}{Nature} \textbf{\bibinfo{volume}{397}},
  \bibinfo{pages}{500} (\bibinfo{year}{1999}{\natexlab{a}}).

\bibitem[{\citenamefont{Lounis and Moerner}(2000)}]{LounisMoerner00}
\bibinfo{author}{\bibfnamefont{B.}~\bibnamefont{Lounis}} \bibnamefont{and}
  \bibinfo{author}{\bibfnamefont{W.}~\bibnamefont{Moerner}},
  \bibinfo{journal}{Nature} \textbf{\bibinfo{volume}{407}},
  \bibinfo{pages}{491} (\bibinfo{year}{2000}).

\bibitem[{\citenamefont{Michler et~al.}(2000)}]{Michler00}
\bibinfo{author}{\bibfnamefont{P.}~\bibnamefont{Michler}} \bibnamefont{et~al.},
  \bibinfo{journal}{Science} \textbf{\bibinfo{volume}{290}},
  \bibinfo{pages}{2282} (\bibinfo{year}{2000}).

\bibitem[{\citenamefont{Moreau et~al.}(2001)}]{Moreau01}
\bibinfo{author}{\bibfnamefont{E.}~\bibnamefont{Moreau}} \bibnamefont{et~al.},
  \bibinfo{journal}{App. Phy. Lett.} \textbf{\bibinfo{volume}{79}},
  \bibinfo{pages}{2865} (\bibinfo{year}{2001}).

\bibitem[{\citenamefont{Beveratos et~al.}(2002)}]{Beveratos02}
\bibinfo{author}{\bibfnamefont{A.}~\bibnamefont{Beveratos}}
  \bibnamefont{et~al.}, \bibinfo{journal}{Euro. Phys. J}
  \textbf{\bibinfo{volume}{18}}, \bibinfo{pages}{191} (\bibinfo{year}{2002}).

\bibitem[{\citenamefont{Yuan et~al.}(2002)}]{Yuan02}
\bibinfo{author}{\bibfnamefont{Z.}~\bibnamefont{Yuan}} \bibnamefont{et~al.},
  \bibinfo{journal}{Science} \textbf{\bibinfo{volume}{295}},
  \bibinfo{pages}{102} (\bibinfo{year}{2002}).

\bibitem[{\citenamefont{Varcoe et~al.}(2001)}]{VarcoeBrattke01}
\bibinfo{author}{\bibfnamefont{B.}~\bibnamefont{Varcoe}} \bibnamefont{et~al.},
  \bibinfo{journal}{Nature} \textbf{\bibinfo{volume}{403}},
  \bibinfo{pages}{743} (\bibinfo{year}{2001}).

\bibitem[{\citenamefont{Brown et~al.}(2003)}]{BrownDani03}
\bibinfo{author}{\bibfnamefont{K.}~\bibnamefont{Brown}} \bibnamefont{et~al.},
  \bibinfo{journal}{Phys. Rev. A} \textbf{\bibinfo{volume}{67}},
  \bibinfo{pages}{043818} (\bibinfo{year}{2003}).

\bibitem[{\citenamefont{Turner et~al.}(1993)}]{TurnerStapelbroek93}
\bibinfo{author}{\bibfnamefont{G.}~\bibnamefont{Turner}} \bibnamefont{et~al.},
  \emph{\bibinfo{title}{Visible light photon counters for scintillating fiber
  applications: I. characteristics and performance}},
  \bibinfo{howpublished}{Proceedings of the Workshop on Scintillating Fiber
  Detectors} (\bibinfo{year}{1993}), \bibinfo{note}{pg. 613}.

\bibitem[{\citenamefont{Takeuchi et~al.}(1999)\citenamefont{Takeuchi, Kim, and
  Yamamoto}}]{TakeuchiKim99}
\bibinfo{author}{\bibfnamefont{S.}~\bibnamefont{Takeuchi}},
  \bibinfo{author}{\bibfnamefont{J.}~\bibnamefont{Kim}}, \bibnamefont{and}
  \bibinfo{author}{\bibfnamefont{Y.}~\bibnamefont{Yamamoto}},
  \bibinfo{journal}{App. Phys. Lett.} \textbf{\bibinfo{volume}{74}},
  \bibinfo{pages}{1063} (\bibinfo{year}{1999}).

\bibitem[{\citenamefont{Atac et~al.}(1994)}]{Atac94}
\bibinfo{author}{\bibfnamefont{M.}~\bibnamefont{Atac}} \bibnamefont{et~al.},
  \bibinfo{journal}{Nucl. Instrum. \& Methods in Phs. Research A}
  \textbf{\bibinfo{volume}{314}}, \bibinfo{pages}{56} (\bibinfo{year}{1994}).

\bibitem[{\citenamefont{Kim et~al.}(1999{\natexlab{b}})\citenamefont{Kim,
  Takeuchi, and Yamamoto}}]{KimTakeuchi99}
\bibinfo{author}{\bibfnamefont{J.}~\bibnamefont{Kim}},
  \bibinfo{author}{\bibfnamefont{S.}~\bibnamefont{Takeuchi}}, \bibnamefont{and}
  \bibinfo{author}{\bibfnamefont{Y.}~\bibnamefont{Yamamoto}},
  \bibinfo{journal}{App. Phys. Lett.} \textbf{\bibinfo{volume}{74}},
  \bibinfo{pages}{902} (\bibinfo{year}{1999}{\natexlab{b}}).

\bibitem[{\citenamefont{Waks et~al.}()}]{WaksInoue03}
\bibinfo{author}{\bibfnamefont{E.}~\bibnamefont{Waks}} \bibnamefont{et~al.},
  \eprint{quant-ph/0308054}.

\bibitem[{\citenamefont{McIntyre}(1966)}]{McIntyre66}
\bibinfo{author}{\bibfnamefont{R.}~\bibnamefont{McIntyre}},
  \bibinfo{journal}{IEEE Trans. Electron Devices}
  \textbf{\bibinfo{volume}{ED-13}}, \bibinfo{pages}{164}
  (\bibinfo{year}{1966}).

\bibitem[{\citenamefont{Kim et~al.}(1997)\citenamefont{Kim, Yamamoto, and
  Hogue}}]{KimYamamoto97}
\bibinfo{author}{\bibfnamefont{J.}~\bibnamefont{Kim}},
  \bibinfo{author}{\bibfnamefont{Y.}~\bibnamefont{Yamamoto}}, \bibnamefont{and}
  \bibinfo{author}{\bibfnamefont{H.}~\bibnamefont{Hogue}},
  \bibinfo{journal}{App. Phys. Lett.} \textbf{\bibinfo{volume}{70}},
  \bibinfo{pages}{2852} (\bibinfo{year}{1997}).

\bibitem[{\citenamefont{Monken et~al.}(1998)\citenamefont{Monken, Ribeiro, and
  Padua}}]{MonkenSoutoRibeiro98}
\bibinfo{author}{\bibfnamefont{C.}~\bibnamefont{Monken}},
  \bibinfo{author}{\bibfnamefont{P.~S.} \bibnamefont{Ribeiro}},
  \bibnamefont{and} \bibinfo{author}{\bibfnamefont{S.}~\bibnamefont{Padua}},
  \bibinfo{journal}{Phys. Rev. A} \textbf{\bibinfo{volume}{57}},
  \bibinfo{pages}{R2267} (\bibinfo{year}{1998}).

\end{thebibliography}

\end{document}